\documentclass[showpacs,aps,superscriptaddress,prb,twocolumn,amsmath]
{revtex4}
\usepackage{txfonts}
\usepackage{color}
\usepackage{amssymb}
\usepackage{graphicx}
\usepackage{dcolumn}
\usepackage{bm}
\usepackage{hyperref}
\usepackage{color}

\begin{document}
\preprint{APS/123-QED}
\title{New Boys and Girls in Phosphorene Family from Gene Recombination: Different from Parents, Excellent than Parents}
\author{Chaoyu He}
\affiliation{Hunan Key Laboratory for Micro-Nano Energy Materials
and Devices, Xiangtan University, Hunan 411105, P. R. China;}
\affiliation{School of Physics and Optoelectronics, Xiangtan
University, Xiangtan 411105, China.}
\author{ChunXiao Zhang}
\affiliation{Hunan Key Laboratory for Micro-Nano Energy Materials
and Devices, Xiangtan University, Hunan 411105, P. R. China;}
\affiliation{School of Physics and Optoelectronics, Xiangtan
University, Xiangtan 411105, China.}
\author{Tao Ouyang}
\affiliation{Hunan Key Laboratory for Micro-Nano Energy Materials
and Devices, Xiangtan University, Hunan 411105, P. R. China;}
\affiliation{School of Physics and Optoelectronics, Xiangtan
University, Xiangtan 411105, China.}
\author{Jin Li}
\affiliation{Hunan Key Laboratory for Micro-Nano Energy Materials
and Devices, Xiangtan University, Hunan 411105, P. R. China;}
\affiliation{School of Physics and Optoelectronics, Xiangtan
University, Xiangtan 411105, China.}
\author{Jianxin Zhong}
\email{jxzhong@xtu.edu.cn}\affiliation{Hunan Key Laboratory for
Micro-Nano Energy Materials and Devices, Xiangtan University, Hunan
411105, P. R. China;} \affiliation{School of Physics and
Optoelectronics, Xiangtan University, Xiangtan 411105, China.}
\date{\today}
\begin{abstract}
Based on the crystal structures of the previously proposed low energy phosphorene allotropes \textbf{$\eta$-P} and \textbf{$\theta$-P} (\textbf{Nano. Lett. 2015, 15, 3557}), we propose five new structural stable phosphorene boys (\textbf{XX-XY} or \textbf{XY-XY}) and girls (\textbf{XX-XX}) through gene (\textbf{XY} from \textbf{$\eta$-P} and \textbf{XX} from \textbf{$\theta$-P}) recombination methods. All of these five new phosphorene allotropes are obviously different from their parents, showing very different and fascinating two-dimensional patterns between each other. The dynamical stabilities of these phosphorene allotropes are confirmed positive and some of them are confirmed energetically more favorable than their parents (\textbf{$\eta$-P} and \textbf{$\theta$-P}). Especially, the \textbf{XX-XX} type girl \textbf{G1-P} is confirmed energetically more favorable than all the previously proposed phosphorene allotropes, including black phosphorene (\textbf{$\alpha$-P, ACS Nano, 2014, 8, 4033}) and blue phosphorene (\textbf{$\beta$-P, Phys. Rev. Lett. 2014, 112, 176802}), which is expected to be synthesized in future experiment through vapor deposition. Our results show that such a new promising phosphorene allotrope \textbf{G1-P} is an excellent candidate for potential applications in nano-electronics according to its middle band gap about 1.58 eV from DFT-HSE06 calculation.\\
\end{abstract}
\maketitle
Black phosphorene (\textbf{$\alpha$-P})\cite{1, 2}, a new atomic thin two-dimensional (2D) material, was successfully exfoliated from its three-dimensional (3D) counterpart black phosphorus through mechanical method at last year. As a new-star in 2D materials family, black phosphorene is considered as a formidable competitor to graphene and other two-dimensional materials for application in nano-electronic fields due to its significant band gap \cite{3, 4} and high carrier mobility\cite{2}. Unfortunately, synthesizing of such a new material with low-cost and high-yielding deposition methods (such as chemical or physical vapor deposition) is still not achieved in the past one year. In contrast, the low pressure amorphous red phosphorus always appears on the substrate in the deposition experiment. That is to say, the possible crystal structure for a 2D phosphorene synthesized from deposition methods is still undetermined, and thus it is still an open question. Theoretically, many possible 2D phosphorene allotropes have been proposed in the past one year, including the black \textbf{$\alpha$-P} \cite{1, 2} (UUUDDD stirrup configuration), blue \textbf{$\beta$-P} \cite{5, 6}(UDUDUD chair configuration), \textbf{$\gamma$-P} \cite{7}(UUDUUD boat-1 configuration), \textbf{$\delta$-P} \cite{7}(UUUUDD boat-2 configuration), \textbf{$\theta$-P} \cite{8}(UUDDUD twist-boat configuration) and red phosphorene \cite{9}(UUUDUD tricycle configuration) in a full 6-6 ring atomic thin layer, the diatomic thin layers \textbf{$\eta$-P} and \textbf{$\theta$-P} with pentagons\cite{10}, as well as some other atomic thin layers with 4-8, 5-7, 5-8 or 3-12 type topological characteristics\cite{7, 10, 11}. In these phosphorene allotropes, the most stable five ones are stirrup black \textbf{$\alpha$-P}, tricycle red phosphorene, diatomic thin \textbf{$\theta$-P}, chair blue \textbf{$\beta$-P} and diatomic thin \textbf{$\eta$-P}, in which that the second stable tricycle red phosphorene is constructed through gene segments recombination based on stirrup black \textbf{$\alpha$-P} and chair blue \textbf{$\beta$-P} in our recent work\cite{9}.\\
\begin{figure*}
\center
\includegraphics[width=6in]{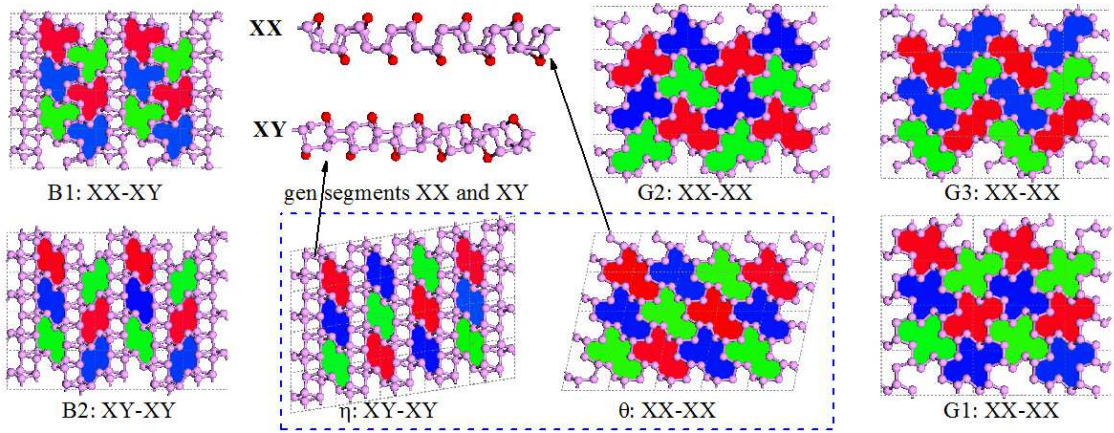}\\
\caption{Top views and corresponding topological patterns of diatomic thin \textbf{$\eta$-P}, \textbf{$\theta$-P} and their descendants \textbf{G1}, \textbf{G2}, \textbf{G3}, \textbf{B1} and \textbf{B2}. The two gene segments \textbf{XX} and \textbf{XY} abstracted form \textbf{$\theta$-P} and \textbf{$\eta$-P}, respectively, are also shown here.)}\label{fig1}
\end{figure*}
\indent Here, we apply the gene segments recombination method to the recently proposed diatomic thin layers \textbf{$\theta$-P} (provide us \textbf{XX} gene segment) and \textbf{$\eta$-P} (provide us \textbf{XY} gene segment) and find five new structural stable phosphorene boys (\textbf{XX-XY} or \textbf{XY-XY}) and girls (\textbf{XX-XX}) with distinct and fascinating 2D topology patterns (See in Fig.1). Density functional theory (DFT) based first-principles method is employed to investigate the structures, energetic stabilities, dynamical stability and electronic properties of these five new possible phosphorene allotropes. Our results show that these five new phosphorene allotropes are dynamically stable and three of them are more favorable than their parents in energy. Especially, the \textbf{XX-XX} type girl \textbf{G1-P} is confirmed energetically more favorable than all the previously proposed phosphorene allotropes, including its mother \textbf{$\theta$-P}, the previously most stable \textbf{$\alpha$-P}, second stable red tricycle phosphorene and fourth stable \textbf{$\beta$-P}. Such a result suggests that \textbf{G1-P} is a promising new 2D material with high probability to be synthesized in future vapor deposition experiments. Our results also show that \textbf{G1-P} is an excellent candidate for potential applications nano-electronics according to its middle band gap of 1.58 eV from DFT-HSE06 calculation. \\
\indent Our calculations of structural optimization and properties investigations are carried out by using the density functional theory (DFT) within generalized gradient approximations (GGA)\cite{12} as implemented in Vienna ab initio simulation package (VASP)\cite{13,14}. The interactions between nucleus and the 3s$^2$3p$^3$ valence electrons of phosphorus atoms are described by the projector augmented wave (PAW) method\cite{15,16}. To ensure the accuracy of our calculations, a plane-wave basis with a cutoff energy of 500 eV is used to expand the wave functions and the Brillouin Zone (BZ) sample meshes are set to be dense enough (less than 0.21 $\AA$$^{-1}$) for each system considered in present work (these settings are set according the convergence test for some important parameters based on the black \textbf{$\alpha$-P} and \textbf{G1-P}). The structures of these five new phosphorene allotropes and some other reference systems considered in present work are fully optimized up to the residual force on every atom less than 0.001 eV/$\AA$. In such a structural optimization process, the optimized exchange van der Waals functional (optB88-vdW)\cite{17,18} is applied to take into account van der Waals interactions. Especially, to gain more reasonable energy band gap of these phosphorene allotropes, the hybrid functional method (HSE06)\cite{19} is considered in the processes of properties investigations after structural optimization. We also simulated the vibrational properties of the five new phosphorene allotropes proposed in our present work through the PHONON package\cite{20} with the forces calculated from VASP to confirm their dynamical stabilities.\\
\begin{figure}
\includegraphics[width=3.5in]{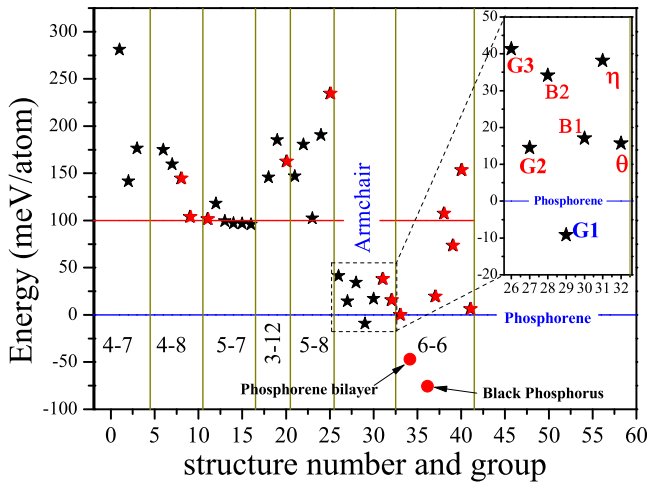}\\
\caption{Total energies per atom of possible phosphorene allotropes are summarized here in different categories. The total energy of \textbf{$\alpha$-P} is set to be zero as reference.}\label{fig2}
\end{figure}
\indent As shown in Fig.1 and Fig.S1, two structural segments can be abstracted from the previously proposed low energy diatomic thin phosphorene allotropes \textbf{$\theta$-P} and \textbf{$\eta$-P}. They are named as \textbf{XY} and \textbf{XX}, respectively, according to their structural characteristics. We can see that the \textbf{XY} (\textbf{XX}) gene is consisted of \textbf{X-Y} (\textbf{X-X}) stacked bilayer of armchair phosphorus chains (and thus we classify \textbf{$\theta$-P}, \textbf{$\eta$-P} and their posterities to category of Armchair) connecting by inter-chain phosphorus atoms (highlighted as red balls). Based on the gene segments \textbf{XX} and \textbf{XY}, we propose five new phosphorene allotropes (In fact, infinite allotropes can be constructed by \textbf{XX} and \textbf{XY} gene segments, but we considered only the situations containing two gene segments per unit cell here) with relatively low energy. We define those containing only \textbf{XX} gene in their bodies as female girls and name them as \textbf{G1}, \textbf{G2} and \textbf{G3}. Those containing \textbf{XY} gene are correspondingly defined as male boys and they are named as \textbf{B1} and \textbf{B2} in our work. In Fig.1, the top views of \textbf{G1}, \textbf{G2}, \textbf{G3}, \textbf{B1}, \textbf{B2} and their parents \textbf{$\theta$-P} and \textbf{$\eta$-P} are shown. We can see that the stacking type between two adjacent \textbf{XX} and/or \textbf{XY} gene segments in each phosphorene allotrope can form individual tiling pattern, which provides us helpful topology characteristics to distinguish them. Only seven staking types are considered in present work, they are \textbf{$\theta$-P}, \textbf{G1}, \textbf{G2} and \textbf{G3} of \textbf{XX-XX}, \textbf{$\eta$-P} and \textbf{B2} of \textbf{XX-XY} and \textbf{B1} of \textbf{XY-XY}. Other stacking types have also been considered in testing but they possess relatively high energy, and thus are not considered in. From Fig.1, we can also see that all these five new allotropes contain two gene segments in their orthorhombic lattice, which are different to their parents those contain only one gene segment in their monoclinic cell. Detail structural information, such as crystalline view from different directions, lattice constants and atomic positions (VASP-POSCAR), of these phosphorene allotropes are prepared in the supplementary file in Fig.S1. One can reproduces these phosphorene crystals according to the optimized POSCARs to further study their structures or investigate their other physical properties.\\
\begin{figure}
\includegraphics[width=3.5in]{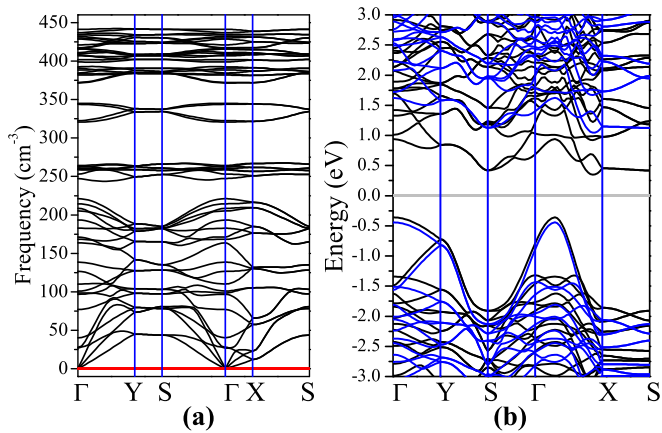}\\
\caption{Phonon band structure (a) and electron band structure (b) of \textbf{G1} calculated form both DFT (black solid line) and HSE06 (blue solid line) methods.}\label{fig3}
\end{figure}
\indent To evaluate the relatively energetic stabilities between these phosphorene allotropes, we calculated their total energies relative to that of the black \textbf{$\alpha$-P}. All previously proposed phosphorene allotropes and some other possible structures (mutations from these previously proposed ones) were also considered in for comparison. The results are classified according to their structural characteristics and plotted in Fig. 2. Our calculations show that the total energy of \textbf{$\theta$-P} and \textbf{$\beta$-P} relative to \textbf{$\alpha$-P} are 15 meV/atom and 19 meV/atom, respectively, which are good consistent with those predicted in previous work\cite{10} and are reliable according to our testing results in cut off energy and K-mesh (See Fig. S2). From Fig.2, we can see that some previously proposed phosphorene allotropes (marked as red five-pointed stars) are not the most stable one in their corresponding categories. For example, in categories of 5-7, 3-12 and 5-8, we find some more favorable candidates (new ones proposed here are marked as black five-pointed stars and please see them in supplementary figures S3-S6). Especially, for the new category of Armchair, the five new allotropes (\textbf{G1}, \textbf{G2}, \textbf{G3}, \textbf{B1} and \textbf{B2}) constructed through gene recombination in our present work show remarkable stability. The total energies of \textbf{G1}, \textbf{G2}, \textbf{G3}, \textbf{B1}, \textbf{B2},  \textbf{$\eta$-P} and \textbf{$\theta$-P} relative to \textbf{$\alpha$-P} are -9 meV/atom, 14 meV/atom, 41 meV/atom, 17 meV/atom, 34 meV/atom, 38 meV/atom and 15 meV/atom, respectively. Such results show that most of these five new phosphorene allotropes are more favorable than their parents and one of them (\textbf{G1}) is more favorable than all the previously proposed 2D phosphorene allotropes including the most stable and experimentally achieved black \textbf{$\alpha$-P}. Although the energy differences between some phosphorene structures are lie in DFT undistinguishable range (level in meV), we still believe that we have found a new phosphorene allotrope \textbf{G1} with excellent stability more favorable than black \textbf{$\alpha$-P}, according our testing results (see in supplementary Fig.S2) and the fact that the bilayer black \textbf{$\alpha$-P} and the 3D black phosphorus are more favorable than single layer black \textbf{$\alpha$-P}. We also believe that there are still many other new forms of phosphorene allotropes will be predicted more favorable than black \textbf{$\alpha$-P} and \textbf{G1} in future. Further theoretical and experimental efforts are expected to be paid on searching for them and synthesizing them.\\
\indent From the view of thermodynamics, low energy generally means high probability to be synthesized in experiments if the system is dynamically possible. The \textbf{XX-XX} type \textbf{G1} with remarkable stability exceeding black \textbf{$\alpha$-P} is expected to be synthesized in future vapor deposition method. We then care about the dynamical stabilities of these new phosphorene allotropes to confirm the possibility of to be synthesized. We evaluate their dynamical stability through simulate their vibrational property. As shown in Fig.3 (a), the phonon band structure of \textbf{G1} phosphorene is free of soft modes associated with structural instabilities. We have also checked the whole Brillouin Zone and find no any imaginary states in its phonon density of states (See Fig.S8). Such results show that allotrope \textbf{G1} is dynamically stable. The dynamical stabilities of the other four new phosphorene allotropes (\textbf{G2}, \textbf{G3}, \textbf{B1} and \textbf{B2}) are also confirmed positive according to their phonon band structures and phonon density of states as shown in supplementary Fig.S7 and Fig.S8. \\
\begin{figure}
\includegraphics[width=3.5in]{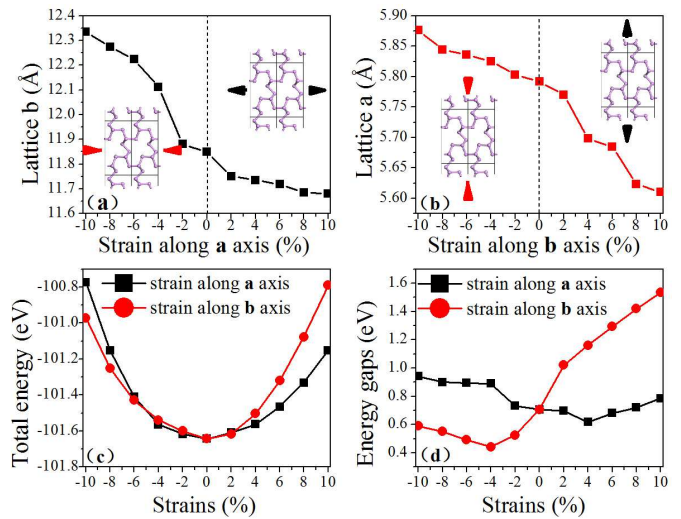}\\
\caption{Strain effects on G1  allotrope, dependence of the lattice b on in-layer stain along direction a (a), dependence of the lattice a on in-layer stain along direction b (b), dependence of total energy on in-layer stains (c) and dependence of the fundamental band gap on in-layer stains (d).}\label{fig4}
\end{figure}
\indent We then care about the fundamental electronic property of such a promising new phosphorene allotrope \textbf{G1}. Band structure of allotrope \textbf{G1} is investigated in both DFT and HSE06 methods. As shown in Fig. 3(b), we can see that allotrope \textbf{G1} is an indirect band gap semiconductor with band gap of 1.58 eV in HSE06 level, which is slightly higher than that of black \textbf{$\alpha$-P}. Our HSE06 calculated band gap for black \textbf{$\alpha$-P} is 1.46 eV, which is good consistent with those reported in previous reports\cite{1}. Such a promising new phosphorene \textbf{G1} with middle band gap is a good candidate for application in nano-electronics. Based on DFT methods, the modulating effect of strains on band gaps of allotrope \textbf{G1} is also investigated. As shown in Fig.4, we can see that allotrope \textbf{G1} show obvious anisotropy. Stains along gen chain (\textbf{XX}) can effectively modulate its band gap and the modulating effect of strain cross gen chains on band gap is not so obvious. The band structures of other allotropes (\textbf{$\eta$-P}, \textbf{$\theta$-P}, \textbf{G2}, \textbf{G3}, \textbf{B1} and \textbf{B2}) are also investigated by both DFT and HSE06 methods and prepared as supplementary in Fig.S9. From these results, we can see that all of these new phosphorene allotropes are indirect band gap semiconductors.\\
\indent In summary, gene segments recombination method is applied to recently proposed diatomic thin layers \textbf{$\eta$-P} and \textbf{$\theta$-P} and five new structural stable phosphorene boys (\textbf{XX-XY} or \textbf{XY-XY}) and girls (\textbf{XX-XX}) with distinct and fascinating 2D topology patterns were proposed to extend phosphorene allotropes family. Our first-principles calculation results show that these five new phosphorene allotropes are dynamically stable and three of them are more favorable than their parents in energy. The \textbf{XX-XX} type girl \textbf{G1-P} is confirmed energetically more favorable than the previously most stable and experimentally achieved black \textbf{$\alpha$-P}. Such a result suggests that \textbf{G1-P} is a promising new 2D material with high probability to be synthesized in future vapor deposition experiments, which is an excellent candidate for potential applications nano-electronics according to its middle band gap of 1.58 eV from DFT-HSE06 calculation.\\
\indent This work is supported by the National Natural Science Foundation of China (Grant Nos. A040204, 11304263 and 11204261), the National Basic Research Program of China (2012CB921303 and 2015CB921103), the Young Scientists Fund of the National Natural Science Foundation of China (Grant No. 11204260), the Scientific Research Found of HuNan Provincial Education department (No. 14C1095), and the Program for Changjiang Scholars and Innovative Research Team in University (IRT13093).\\


\begin{thebibliography}{20}
\bibitem{1} H. Liu, A.T. Neal, Z. Zhu, Z. Luo, X. Xu, D. Tom\'{a}nek, and P.D. Ye,  ACS Nano, \textbf{8}, 4033 (2014).
\bibitem{2} L. Li, Y. Yu, G.J. Ye, Q. Ge, X. Ou, H. Wu, D. Feng, X.H. Chen, and Y. Zhang, Nat. Nano, \textbf{5}, 372 (2014).
\bibitem{3} Y. Liu, F. Xu, Z. Zhang, E.S. Penev, and B.I. Yakobson, Nano lett., \textbf{14}, 6782 (2014).
\bibitem{4} L. Liang, J. Wang, W. Lin, B.G. Sumpter, V. Meunier, and M. Pan, Nano Lett., \textbf{14}, 6400 (2014).
\bibitem{5} Z. Zhu, and D. Tom\'{a}nek, Phys. Rev. Lett., \textbf{112}, 176802 (2014).
\bibitem{6} S.E. Boulfelfel, G. Seifert, Y. Grin, and S. Leoni, Phys. Rev. B, \textbf{85}, 014110 (2012).
\bibitem{7} J. Guan, Z. Zhu, and D. Tom\'{a}nek, Phys. Rev. Lett., \textbf{113}, 046804 (2014).
\bibitem{8} J. Guan, Z. Zhu, and D. Tom\'{a}nek, ACS Nano., \textbf{8}, 12763 (2014).
\bibitem{9} T. Zhao, C.Y. He, S.Y. Ma, K.W. Zhang, X.Y. Peng, G.F. Xie, and J.X. Zhong, J. Phys.: Condens. Matter., \textbf{27}, 265301 (2015).
\bibitem{10} M.H. Wu, H.H. Fu, L. Zhou, K.L. Yao, and X.C. Zeng, Nano lett., \textbf{15}, 3557 (2015).
\bibitem{11} G.D. Yu, L.W. Jiang, and Y.S. Zheng, J. Phys.: Condens. Matter., \textbf{27}, 255006 (2015).
\bibitem{12} J.P. Perdew, and W. Yue., Phys. Rev. B, \textbf{33}, 8800 (1986).
\bibitem{13} G. Kresse and J. Furthm\"{u}ller, Phys. Rev. B, \textbf{54}, 11169 (1996).
\bibitem{14} G. Kresse and J. Furthm\"{u}ller, Comput. Mater. Sci., \textbf{6}, 15 (1996).
\bibitem{15} P. E. Bl\"{o}chl, Phys. Rev. B \textbf{50}, 17953 (1994).
\bibitem{16} G. Kresse and D. Joubert, Phys. Rev. B \textbf{59}, 1758 (1999).
\bibitem{17} J. Klimes, D. R. Bowler, and A. Michaelides, J. Phys.: Cond. Matt., \textbf{22}, 022201 (2010).
\bibitem{18} J. Klimes, D. R. Bowler, and A. Michaelides, Phys. Rev. B, \textbf{83}, 195131 (2011).
\bibitem{19} J. Heyd, G. E. Scuseria, and M. Ernzerhof, J. Chem. Phys. \textbf{118}, 8207 (2003).
\bibitem{20} K. Parlinski, Z-. Q. Li, and Y. Kawazoe, Phys. Rev. Lett. \textbf{78}, 4063 (1997).
\end{thebibliography}
\end{document}